\documentclass[twocolumn, aps, pra, showpacs, superscriptaddress, floatfix, nofootinbib, 10pt]{revtex4-2} 

\usepackage{amsmath,bbm,amsthm}
\usepackage{amssymb}
\usepackage{graphicx}
\usepackage{color}
\usepackage{tcolorbox}
\usepackage[colorinlistoftodos]{todonotes}
\usepackage[colorlinks=true, allcolors=blue]{hyperref}
\usepackage{dsfont}
\usepackage{float}
\usepackage{cancel}
\usepackage[autostyle]{csquotes}
\usepackage{enumitem}   
\usepackage{mathrsfs}
\usepackage[T1]{fontenc}
\usepackage{soul}

\newtheorem{theorem}{Theorem}

\newtheorem{proposition}{Proposition}

\newtheorem*{theorem*}{Theorem}

\newcommand{\cG}{\mathcal{G}}
\newcommand{\cK}{\mathcal{K}}
\newcommand{\cN}{\mathcal{N}}

\newcommand{\match}{\mathrm{color}}
\newcommand{\yellow}{\mathrm{yellow}}
\newcommand{\purple}{\mathrm{purple}}
\newcommand{\red}{\mathrm{red}}

\newcommand{\beq}{\begin{equation}}
\newcommand{\eeq}{\end{equation}}
\newcommand{\beqa}{\begin{eqnarray}}
\newcommand{\eeqa}{\end{eqnarray}}

\newcommand{\ket}[1]{\ensuremath{\left|#1\right\rangle}}

\def\opone{\leavevmode\hbox{\small1\normalsize\kern-.33em1}}

\newcommand{\dd}{\mathrm{d}}

\renewcommand{\today}{\number\day\space\ifcase\month\or
   January\or February\or March\or April\or May\or June\or
   July\or August\or September\or October\or November\or December\fi
   \space\number\year}

\begin{document}

\title{Nonlocality for Generic Networks}
\date{\today}
\author{Marc-Olivier Renou}
\affiliation{ICFO-Institut de Ciencies Fotoniques, 08860 Castelldefels (Barcelona), Spain}
\author{Salman Beigi}
\affiliation{School of Mathematics, Institute for Research in Fundamental Sciences (IPM), P.O.~Box 19395-5746, Tehran, Iran}

\begin{abstract}
Bell's theorem shows that correlations created by a single entangled quantum state cannot be reproduced classically.
Such correlations are called \emph{Nonlocal}.
They are the elementary manifestation of a broader phenomenon called \emph{Network Nonlocality}, where several entangled states shared in a network create \emph{Network Nonlocal} correlations.
In this paper, we provide the first class of strategies producing nonlocal correlations in generic networks.
In these strategies, called \emph{Color-Matching} (CM), any source takes a color at random or in superposition, where the colors are labels for a basis of the associated Hilbert space.  
A party (besides other things) checks if the color of neighboring sources match. 
We show that in a large class of networks without input, well-chosen quantum CM strategies result in nonlocal correlations that cannot be produced classically.
For our construction, we introduce the graph theoretical concept of rigidity of classical strategies in networks, and using the Finner inequality, establish a deep connection between network nonlocality and graph theory. In particular, we establish a link between CM strategies and the graph coloring problem. This work is extended in a longer paper~\cite{PRA}, where we introduce a second family of rigid strategies called Token-Counting, leading to network nonlocality. 
\end{abstract}

\maketitle

Bell's theorem demonstrates the Nonlocal behavior of quantum correlations arising from local measurements of an entangled state~\cite{Bell1964}, a fundamental signature of quantum physics. 
It relies on the Bell inequalities, which provide us with standard methods to characterize nonlocal correlations of a single quantum state~\cite{BrunnerRMP}. 
These methods constitute a powerful toolbox for a large spectrum of applications, particularly for the black box certification of states and measurements~\cite{Mayer1998QuantCryptoUnsecAppar,Supic2020SelfTestReview,Kaniewski2016SelfTest} that is crucial in the study of random number generators~\cite{Pironio2010DIQRNG, Herrero2019} and quantum cryptographic protocols~\cite{Bennett1984BB84, Acin2007DIQKD_CollAttacks}.
 To cope with technical challenges, nowadays experiments often involve networks of several sources used to simulate standard single-source Bell protocols. 
This is for instance the case in the first loophole-free violation of a Bell inequality~\cite{HensenLoopholeFree2015}, where the EPR pair used for the violation of the CHSH inequality is created through entanglement swapping of two distinct sources. 
Similar procedures are envisioned for long range QKD protocols using multimode quantum repeaters to cope with loss noise in optical fibers~\cite{Sangouard2011QuantRepeaters}. 

In a significant conceptual advance, it was recently understood that networks are much more versatile than a tool to simulate a single quantum source. 
Beyond this simple use, they can lead to new forms of nonlocality called \emph{Network Nonlocality}, which is imposed by the network topology.
This remarkable point of view can be exploited for new applications such as certification of entangled measurements~\cite{Renou2018BSMSelfTest,Bancal2018BSMSelfTest}, or more surprisingly, detection of the nonlocality of \emph{all} entangled states~\cite{Bowles2018DIAllEntState}, which is out of reach in standard Bell scenarios~\cite{Tsirelson1993}. 
At a conceptual level, it was recently used to justify the misdefinition of the notion of genuine multipartite entanglement~\cite{navascues2020gnme,kraft2020networkentanglement,luo2020networkentanglement} and genuine multipartite nonlocality~\cite{Coiteux2021a,Coiteux2021b}, and understand the role of complex numbers in quantum theory~\cite{Renou2021}. 

More fundamentally, several techniques were developed to characterize classical (also called local) and nonlocal correlations in networks \cite{wolfe2019inflation,wolfe2021qinflation,Pozas2019} and various examples of quantum Network Nonlocal correlations are known.
For instance, Fritz has shown that any standard Bell nonlocality scenario can be embedded in networks~\cite{fritz2012}, resulting in various network nonlocal correlations. 
In particular, Fritz's work proves that inputs are not necessary to obtain network nonlocal correlations.
The first example of Network Nonlocality, obtained via entanglement swapping in the bilocal network~\cite{branciard2010,branciard2012,gisin2017}, was generalized to star networks~\cite{tavakoli2017} and single-node cluster networks~\cite{luo2018}. 
These examples are all based on the extension of the CHSH inequality~\cite{rosset2016}. 
Recently, another example in the bilocal network was derived~\cite{tavakoli2020} that is different in nature from the previous ones.   
Also, a new form of Network Nonlocality, qualified of genuine Network Nonlocality~\footnote{
We emphasize that the notion of genuine Network Nonlocality differs from the one of genuine multipartite entanglement/multipartite  nonlocality. The later designs entanglement or nonlocality which can only be obtained from a single global shared nonclassical source (and not, for instance, in a network). Genuine Network Nonlocality designs a form of nonlocality which ``genuinely'' comes from the network structure, and not from the standard Bell nonlocality arguments.
}, was found in the triangle network with no input~\cite{Renou2019a}.
Nevertheless, except for the last example that is also generalized to ring networks with an odd number of nodes, {
no construction of network nonlocality in generic networks,
}
 unrelated to the standard Bell nonlocality, is known.

In this paper, we provide the first method to obtain examples of network nonlocality in a wide class of networks with no input, unrelated to Fritz's embedding of Bell's nonlocality. 
Our method is based on the family of \emph{quantum Color-Matching} (CM) strategies, in which each source distributes a coherent superposition of some colors,
where by colors we mean labels for the elements of a canonical basis of the Hilbert space of each source.
{
Each party applies a measurement to output either a matching color (meaning that the color of all the connected sources match) or some refined information in the case of no color match. This results in a joint probability distribution $P$ over the outputs of the parties.  Importantly, keeping only the color part of the outputs contained in a coarse grained distribution $P_\match$ of $P$, we show that in a large class of networks, $P_\match$ is \emph{rigid}, meaning that there is essentially a unique classical strategy to simulate $P_\match$.
Based on this rigidity property, we obtain network nonlocal correlations in a large family of networks.
}

\begin{figure}
\includegraphics[width=2.5in]{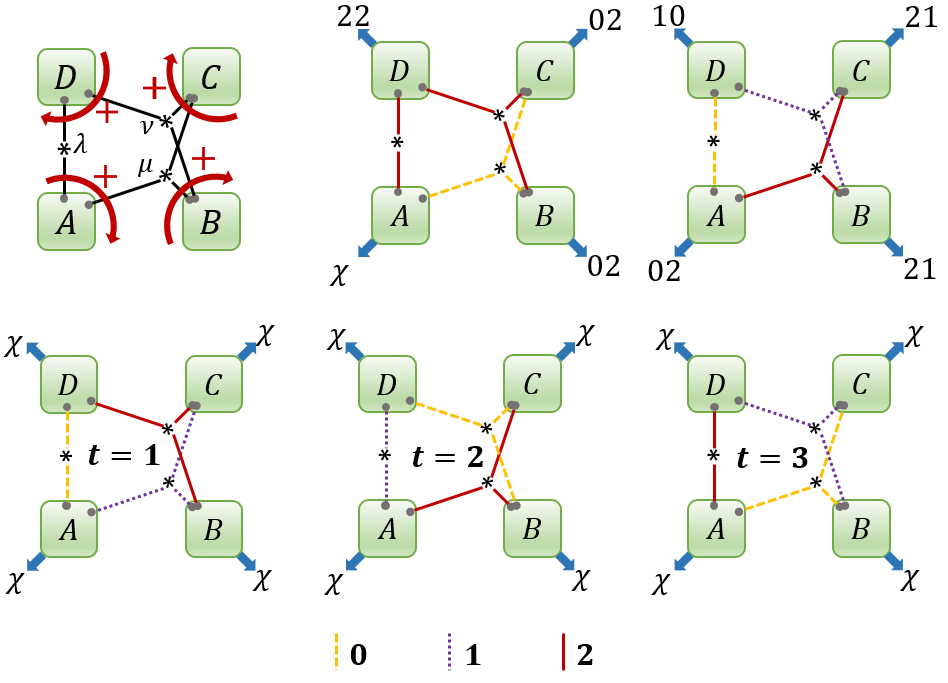}
\caption{Four-party 1-2 network, source orientation and illustration of some events of the decohered classical CM strategy leading to $P_\match$. All outputs except the ambiguous $\chi$ give the color of the sources adjacent to a party. Hence, there exist only three patterns for the color of the sources for which all outputs are ambiguous, that are labelled by $t=1, 2, 3$.
}\label{fig:1-2CM}
\end{figure}

\begin{figure}
\center
\includegraphics[width=2.5in]{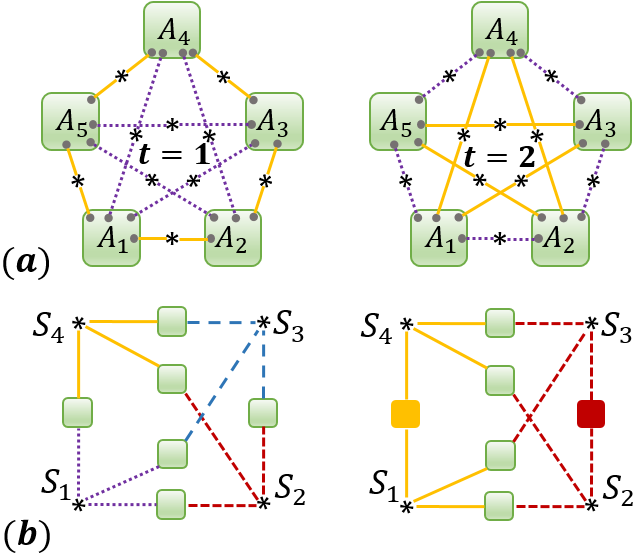}
\caption{$(a)$ Network nonlocality in Bipartite Source Complete networks  $\cK_{n}$. The network is composed of $n$ parties and $m=\binom{n}{2}$ bipartite sources, such that each pair of parties share a source (here $n=5, m=10$). In~\cite{PRA}, we introduce a quantum CM strategy with two colors
 in which $\mathcal T$ contains all tuples of colors which do not correspond to the cases $t=1$ and $t=2$ in the figure. 
Theorem~\ref{thm:RigidityCMGeneral} applied to the decohered classical strategy implies that any hypothetical classical simulation of the resulting distribution must assign colors to sources. Hence, assuming that all parties' outputs are ambiguous, the unobserved classical hidden variable $t$ is well-defined. This introduces new constraints, forming a linear program, which has no solution.
\\
$(b)$ Network nonlocality in Bipartite Parties Complete networks  $\cG_{m}$. The network is composed of $m$ sources and $n=\binom{m}{2}$ parties, such that every pair of sources have one common party (here $m=4, n=6$). In~\cite{PRA}, we introduce a quantum CM strategy with $m$ colors. In this strategy, based on the graph coloring problem, each party checks if the color of adjacent sources match. Thus, when all parties find a non-colored output, the underlying graph (with parties being the edges) is properly colored (no adjacent sources have the same color). For example the {left colored network} is properly colored, while {in the right colored network} two parties with their red and yellow outputs can detect the non-proper coloring.
As before a well chosen set $\mathcal T$ allows us to prove nonlocality.
}\label{fig:BipSComplNet_BipPComplNet}
\end{figure}

We first illustrate our construction of network nonlocality in the simple four-party scenario of Fig.~\ref{fig:1-2CM}.
Then, we present our method in full generality, demonstrating network nonlocality

 in bipartite source complete networks $\cK_n$ and in bipartite party complete networks $\cG_m$ (see Fig.~\ref{fig:BipSComplNet_BipPComplNet}). This last example demonstrates a connection between our CM distributions and the graph coloring problem.  
The ideas presented in this letter are further developed in~\cite{PRA}, where we introduce a second family of rigid strategies, called Token-Counting, in which each source distributes a fixed number of tokens to its adjacent parties, and parties count the number of received tokens.
 
\medskip

\textit{First example of nonlocality from Color-Matching---}
Consider the  four-party scenario of Fig.~\ref{fig:1-2CM} and the following quantum Color-Matching (CM) strategy on this network:
\begin{enumerate}[itemsep=-1mm, topsep=0pt]
\item The sources distribute a coherent uniform superposition of the colors $0~(\yellow)$, $1~(\purple)$ or $2~(\red)$, i.e., source $\lambda$ distributes $\frac{1}{\sqrt{3}}(\ket{00}+\ket{11}+\ket{22})$ and sources $\mu, \nu$ distribute $\frac{1}{\sqrt{3}}(\ket{000}+\ket{111}+\ket{222})$.
\item Each party receives a pair of colors ordered as in Fig.~\ref{fig:1-2CM}, and checks if this ordered pair is one of $00, 11, 22, 02, 21$ or $10$.  
In other words,  each party first applies a measurement consisting of projections on each of vectors {$\ket{00}, \ket{11}, \ket{22}, \ket{02}, \ket{21}, \ket{10}$} and a projection, denote by $\Pi_\chi$, on the orthogonal complement of these vectors spanned by $\{\ket{01}, \ket{12}, \ket{20}\}$. 
The three projectors on basis vectors $\ket{00}, \ket{11}, \ket{22}$ correspond to what we call color matches. 

\item If a party's measurement outcome is $\Pi_{\chi}$ (i.e., not one of $00, 11, 22, 02, 21, 10$), then she performs a refined measurement.
{That it, she measures in the basis $\ket{\chi_{r}}=\alpha_r\ket{01}+\beta_r\ket{12}+\gamma_r\ket{20}$, ${r=1,2,3}$, where $\alpha_1, ..., \gamma_3\in\mathds{C}$ are parameters that satisfy orthonormalization constraints.}
We note that the states $\ket{\chi_{r}}$ for $r=1, 2, 3$ span the same space as the support of $\Pi_\chi$.
\end{enumerate}

{We let $P=\{P(a, b, c, d)\}$ be the \emph{CM quantum distribution} resulting from this strategy, where $a, b, c, d\in\{00, 11, 22, 02, 21,10, \chi_{1}, \chi_{2}, \chi_{3}\}$ are the parties' outputs and the probabilities $P(a, b, c, d)$ can be computed with the Born rule.
We also introduce a coarse graining $P_\match$ of $P$, which is obtain when the parties ignore the exact value of $r$ in $\ket{\chi_r}$, and replace these three outputs by a unique output $\chi$: in $P_\match=\{P_\match(a, b, c, d)\}$, the outputs belong to $\{00,11, 22, 02, 21,10, \chi\}.$ Equivalently, $P_\match$ is obtained when the parties ignore step 3 of the protocol above, and measure with the projectors over the states $\ket{00}, \ket{11}, \ket{22}, \ket{21}, \ket{10}, \ket{02}$ and the rank-three projection $\Pi_\chi$.
}

We are interested in the possibility of simulating $P$ by a \emph{classical strategy}.
In a classical strategy, $\lambda, \mu, \nu$ are independent sources distributing classical randomness, and the parties' outputs are some probabilistic functions of the values taken by their adjacent sources{, denoted by $P(a|\lambda,\mu), P(b|\mu,\nu), P(c|\mu,\nu), P(d|\nu,\lambda)$} .
{For such a classical strategy, the distribution is given by
\begin{align}
P(a,b,c,d)&=\nonumber\\
\int \dd\lambda\dd\mu&\dd\nu P(a|\lambda,\mu) P(b|\mu,\nu) P(c|\mu,\nu) P(d|\nu,\lambda).\label{eq:ClassicalModel}
\end{align}}
If there is no classical strategy that simulates $P${, i.e., if there does not exist any probabilistic output functions allowing to obtain $P$ as in~\eqref{eq:ClassicalModel},} we say that $P$ is \emph{network nonlocal}.

The first main result of this letter is the following proposition, which will be generalized to a method that can be applied to a large variety of networks:
\begin{proposition}\label{propo:CM1-2}
For some well-chosen parameters $\alpha_1, ..., \gamma_3\in\mathds{C}$ satisfying the orthonormalization constraints, the distribution $P$ is network nonlocal.
\end{proposition}

{Note that a necessary (but a priori not sufficient) condition to obtain a network nonlocal distribution $P$ is that some $\ket{\chi_{r}}$ are entangled.}

{We sketch the proof of this proposition below, in three steps, and leave a complete proof for~\cite{PRA} in which we exhibit particular choices of parameters achieving a non-local distribution $P$. }

\begin{proof}
\emph{First step: classical CM strategy for $P_\match$}.
{In this step, we} observe that the measurement operators associated to $P_{\match}$, including $\Pi_\chi$, are all diagonal in the computational basis. 
Hence, the output distribution $P_{\match}$ does not change if the source's colors are measured in the computational basis before that they are sent to the parties. 
As this computational basis measurement demolishes the coherence of sources, the resulting strategy is classical, and we call it the associated \emph{decohered classical CM strategy}. 
In this strategy, each source independently takes one of the colors $0$, $1$ or $2$ with equal probability $1/3$ (in a non-coherent classical way), and each party declares when the received color-pair belongs to $\{00, 11, 22, 02, 21, 10\}$, and if not, declares the ambiguous output $\chi$.
Remark that there are only three patterns for the colors of sources for which
\emph{all} outputs are ambiguous. 
These three patterns are depicted in Fig.~\ref{fig:1-2CM} by $t=1,2,3$.

\emph{Second step: Rigidity of $P_\match$.}
{
In this step, we show that up to local relabelling of the colors and addition of irrelevant information to the sources, the decohered classical CM strategy is the unique classical strategy for simulating $P_\match$. 
More precisely, we prove that for source (say) $\lambda$ there is a color function $c_\lambda$ that assigns a color in $\{0, 1, 2\}$ to any value  that might be taken by source $\lambda$ in that classical strategy. 
 Moreover, to determine her output, a party looks only at this color, i.e., $c_\lambda$, and not at the exact value taken by the sources. 
Finally, we prove that $c_\lambda$ takes each color $0, 1, 2$ with probability $1/3$, similar to that of the decohered classical CM strategy. 
We call this the rigidity property of $P_\match$. Our proof of this step is detailed in~\cite{PRA} and sketched below in Theorem~\ref{thm:RigidityCMGeneral}, which generalizes this rigidity notion.
}%
%
~\\
\emph{Third step: Impossibility of the extension of the classical CM strategy.}
{
In this last step, we assume the existence of a strategy that simulates $P$ and obtain a contradiction to finish the proof.
Consider such a hypothetical simulating classical strategy. We first remark that this strategy can be used to simulate $P_\match$ by coarse-graining. This allows to apply the rigidity property demonstrated in step 2: there exists functions $c_\lambda, c_\mu, c_\nu$ that associate well-defined color in $\{0, 1, 2\}$ to all values of the respective classical sources.
}
 Now consider the case where all parties' outputs are ambiguous{, i.e., $(a, b, c, d)=(\chi_{i},\chi_{j},\chi_{k},\chi_{l})$} for some $1\leq i, j, k, l\leq 3$. This happens only if the colors taken by the sources follow one of the three patterns of Fig.~\ref{fig:1-2CM} indicated by $t=1, 2, 3$. 
{
Hence, when all parties obtain an ambiguous output, we can introduce an unobserved classical \emph{hidden variable} $t\in\{1,2,3\}$.
}
We show in~\cite{PRA} that the existence of $t$ imposes some new constraints on this hypothetical classical strategy.
{Indeed, it imposes the existence of a joint distribution  $q(a=\chi_{i},b=\chi_{j},c=\chi_{k},d=\chi_{l},t)$ some of whose marginals can be explicitly computed in terms of the parameters $\alpha_1, \dots,  \gamma_3$. }
{These marginals put constraints on the joint distribution $q(a=\chi_{i},b=\chi_{j},c=\chi_{k},d=\chi_{l},t)$ which  form a linear program. This linear program can be shown to be infeasible for well-chosen parameters $\alpha_1, \dots,  \gamma_3$,  refuting the existence of the joint distribution and the hidden variable $t$. 
Thus, we conclude that no classical strategy can simulate $P$. }
\end{proof}

\textit{Nonlocality for generic networks---}
{We now consider an arbitrary network $\cN$ with sources $S_1, \dots S_m$ and parties $A_1, \dots,  A_n$.
We sketch a generalization of the above example to construct nonlocal correlations in a large class of networks $\cN$,  with details being in~\cite{PRA}. 
Our generalization requires that $\cN$ is an Exclusive Common-Source (ECS) network, i.e., for any of sources $S_i$ of $\cN$, there exist $A_j\neq A_{j'}$ such that $S_i$ is the only source connected to both $A_j$ and $A_{j'}$. 
Our technical proof also requires that $\cN$ admits a Perfect Fractional Independent Set (PFIS). 
As it is unclear whether the latter condition is really necessary for our construction, or there exists an alternative proof avoiding it,\footnote{{For some ECS networks not admitting a PFIS, CM rigidity still holds (private communication with Sadra Boreiri).}} we give the definition of PFIS only in~\cite{PRA}. We note, however, that the network of Fig.~\ref{fig:1-2CM} as well as all \emph{regular} networks in which all sources are connected to the same number of parties admit a PFIS.} 

{Let $\cN$ be an ECS network admitting a PFIS.} We fix a set of colors $\{1, \dots,  C\}$
. 
We introduce the following \emph{classical CM strategy}:

\begin{enumerate}[itemsep=-1mm, topsep=0pt]
\item Each source $S_j$ {independently and uniformly} picks a color $c_j\in \{1, \dots,  C\}$.
\item Each party $A_j$ checks if the colors she observes from connected sources are compatible with some predetermined set of tuples of colors denoted by $\mathcal T$. Importantly, we assume that $\mathcal T$ contains all tuples of the form $(c, \dots, c)$ for some color $c$, which means that if a party receives the same color from all its adjacent vertices, she outputs the color match $c$. 
\item If the observed colors are not compatible with any of the tuples contained in $\mathcal T$, then $A_j$ outputs an ambiguous symbol $\chi$. 
\end{enumerate}
We denote by $P_\match$ the \emph{CM classical distribution} resulting from this strategy.
Remark that this definition generalizes our decohered classical CM strategy for the network of Fig.~\ref{fig:1-2CM}. {In this example, the set $\mathcal T$ equals $\{ 000, 111, 222, 021, 210, 102\}$, and if a party's received colors are compatible with an element of $\mathcal T$, she outputs the received colors; otherwise she outputs $\chi$.\footnote{{Note that a party has not access to all colors. E.g., when $A$ sees $c_\lambda=0, c_\mu=2$, she outputs $a=02$ as what she sees is compatible with the tuple $(c_\lambda, c_\mu, c_\nu)=(0, 2, 1)\in \mathcal T$; whereas when she sees $c_\lambda=0, c_\mu=2$, she outputs $a=\chi$ as what she sees is not compatible with any tuple $(c_\lambda,  c_\mu, c_\nu)\in \mathcal T$}.}}
The following theorem states the rigidity of $P_{\match}$ in full generality.
\begin{theorem}[Rigidity of $P_\match$, general]\label{thm:RigidityCMGeneral}
For any ECS network $\cN$ admitting a PFIS, the distribution $P_\match$ is rigid.
This means that, for any classical strategy (not necessary CM) that simulates $P_\match$ in $\cN$, there exist {uniformly distributed source color functions} $c_i(s_i)\in\{1,...,C\}$ associating a color $c_i(s_i)$ to each value $s_i$ taken by a source $S_i$, such that each party looks only at the color functions to determine her input (and not to the exact values of $s_i$'s). 
\end{theorem}

\begin{proof}[Proof sketch.]
{
In~\cite{PRA}, we consider the Finner inequality~\cite{Finner1992,RenouFinner2019}, which is a generalization of the H\"older inequality constraining the potential distributions that can be classically simulated in a network.  To write down the Finner inequality we need some weights which we assume are taken from the granted PFIS. With these weights we realize that $P_{color}$ satisfies the Finner inequality with an equality condition. Thus, we explore the equality condition of Finner's inequality. Using these conditions, we demonstrate that for any classical strategy that simulates $P_\match$, there must exist an assignment of colors to the values taken by sources.
At last, thanks to the ECS assumption, we show that these functions are consistent and satisfy all the desired properties. 
}
\end{proof}

{Note that this theorem can be thought of as a self-testing of a classical strategy in a network: only looking at the parties' outputs, one is able to understand  the internal functioning of sources and parties in a classical strategy for $P_\match$.}
It is the starting point of a method to obtain examples of nonlocality in various networks. In this method, we consider a quantum CM strategy in which sources distribute colors in superposition and parties' measurement bases consist of two types of projectors:  projections that are diagonal in the color basis and projections that are not diagonal in this basis. The diagonal projectors correspond to color-tuples compatible with the tuples in the set $\mathcal T$  introduced above, and in particular contain color matches. This gives the quantum CM distribution $P$.
The nonlocality follows the method used in our first example step by step.

We provide more details on our method in~\cite{PRA}, where we apply it to various networks. In particular, considering two colors, we show that all bipartite source complete networks $\cK_n$ of $n$ parties (where any pair of parties $A_j,A_{j'}$ share a bipartite source $S_{jj'}$) admit network nonlocality.
We also consider the class of bipartite party complete networks $\cG_m$ consisting of $m$ sources where any pair of sources $S_i, S_{i'}$  are connected to a party $A_{ii'}$ (who is not connected to any other source). Introducing a method based on graph coloring, we prove network nonlocality for $\cG_m$ as well. See Fig.~\ref{fig:BipSComplNet_BipPComplNet} for more details on these two examples.

\textit{Discussion---}
In this paper we proposed a general method for deriving nonlocality in a wide class of networks. Our method is based on the crucial observation that Color-Matching distributions are rigid. That is, in order to classically simulate such distributions (in certain networks) we are forced to use CM strategies. This rigidity property substantially restricts the range of  classical strategies that can simulate such distributions. Then, further study of these strategies leads us to examples of  nonlocality in networks.

As argued by Fritz~\cite{fritz2012}, we may embed nonlocal distributions of standard Bell's scenarios in networks. For instance, a standard quantum violation of the CHSH inequality can be turned into examples of nonlocality in the triangle network by interpreting the two inputs as two new independent sources.
 This embedding,
{
in which several sources and parties of the network only have a classical behavior,
}
can be generalized for a large class of networks. Nevertheless, in~\cite{PRA}, we argue that our examples of network nonlocality are fundamentally different from this embedding of Bell's nonlocality into network scenarios, as our examples imply the creation of 
{
a global entangled state involving all sources and parties of the networks.
}

Let us now discuss a potential experimental implementation. Our examples use maximally entangled states at the sources
{
, yet we prove in~\cite{PRA} that our method allows to consider any possible {(non maximally) entangled state}, offering a large adaptability to experimental capabilities.
Also, in~\cite{PRA}, we explain how a noise tolerant version of our result can be obtained. 
Hence, our result still holds for sources distributing noisy states. 
More precisely, the rigidity result of Theorem~\ref{thm:RigidityCMGeneral} holds even for disturbed CM distributions. The point is that the second step of the proof of this theorem adapts to the noisy regime, where an approximate equality condition in the Finner inequality~\cite{EFKY16} can be applied to show the existence of \emph{noisy color functions}. These noisy functions can be used to prove the network nonlocality of concrete imperfect implementations of, say, our first example in the network of Fig.~\ref{fig:1-2CM}.
However, this direct adaptation of our proof results in an extremely weak (experimentally not realistic) noise tolerance.
It would be desirable to find new proof techniques for the rigidity of CM distributions that are well-adapted in the noisy regime.
Alternatively, one may consider optimization approaches to estimate this noise tolerance. 
For instance, the recent machine learning algorithms developed in~\cite{Krivvachy2020} already predicted an experimentally more reasonable noise tolerance for the network nonlocal distribution of~\cite{Renou2019a}.}\footnote{{Note that the noise tolerance deduced from an optimization algorithm such as the one developed in~\cite{Krivvachy2020} cannot be considered as a rigorously proven noise tolerance value, as these algorithms only do a non-certified optimization over the potential models of the form of Eq.~\eqref{eq:ClassicalModel}. However, for a concrete proof of concept in an experiment which would anyway involve other experimental assumptions not rigorously provable, it might be enough to rely on our rigorous proof for the ideal case (and extremely weak noisy case), and on the estimated noise tolerance.  This would require a high level of confidence in the optimization algorithm used: such discussions are out of the scope of our letter.}} 
{This work provides a clear indication that there should exist a strong enough noise tolerance on our example allowing for experimental verifications.
}
%
%
%

%
Finally, we consider CM strategies as one example of a potential general method to derive Network Nonlocality based on graph theoretical tools. 
In~\cite{PRA}, we show that there exists another family of strategies, which we call Token-Counting (TC), that satisfy a rigidity property and lead to examples of nonlocality in a large class of networks.
This allows to reinterpret our previous work~\cite{Renou2019a} as the simplest example of a TC strategy~\cite{TavakoliReview,Abiuso2021}. 
Far from being an isolated case of network nonlocality, our extended work~\cite{PRA} shows that~\cite{Renou2019a} is the fingerprint of a rigidity-based method revealing the nonlocality in a large class networks. 
The present letter shows that beyond TC strategies, CM strategies is another family of rigid strategies allowing to exhibit nonlocal distributions in ECS networks admitting a PFIS, a totally different class of networks.
Despite the fact that they are based on fundamentally different proof techniques, both CM and TC strategies can be used to reveal nonlocality from their rigidity properties. 
We conjecture that they are parts of a general method to derive Network Nonlocality based on combinatorial and graph theoretic primitives.

\medskip
\textit{Acknowledgements---}
We thank Antonio Ac\'in and Nicolas Gisin for discussions.
M.-O.R. is supported by the Swiss National Fund Early Mobility Grants P2GEP2\_19144 and the grant PCI2021-122022-2B financed by MCIN/AEI/10.13039/501100011033 and by the European Union NextGenerationEU/PRTR, and acknowledges the Government of Spain (FIS2020-TRANQI and Severo Ochoa CEX2019-000910-S [MCIN/ AEI/10.13039/501100011033]), Fundació Cellex, Fundació Mir-Puig, Generalitat de Catalunya (CERCA, AGAUR SGR 1381) and the ERC AdG CERQUTE.

\bibliographystyle{apsrev4-2}
\bibliography{references}

\end{document}